\documentclass[preprint,12pt]{elsarticle}

\usepackage{graphics}

\usepackage{amssymb}
\usepackage{latexsym}

\journal{Physica A}

\begin{document}

\begin{frontmatter}



\title{
Critical behavior of the fidelity susceptibility
for the $d=2$ transverse-field Ising model
}


\author{Yoshihiro Nishiyama} 

\address{Department of Physics, Faculty of Science,
Okayama University, Okayama 700-8530, Japan}

\begin{abstract}
The overlap (inner product) between the ground-state
eigenvectors
with proximate interaction parameters,
the so-called fidelity,
plays a significant role in 
the quantum-information theory.
In this paper, the critical behavior 
of the fidelity susceptibility is investigated for
the two-dimensional tranverse-field (quantum) Ising model
by means of the numerical diagonalization method.
In order to treat a variety of system sizes
$N=12,14,\dots,32$,
we
adopt the screw-boundary condition.
Finite-size artifacts (scaling corrections) 
of the fidelity susceptibility
appear to be suppressed, 
as compared to those of the Binder parameter.
As a result, we 
estimate the fidelity-susceptibility critical exponent
as $\alpha_F =0.715(20)$.
\end{abstract}


\end{frontmatter}



\section{\label{section1}Introduction}

In the quantum-information theory,
the inner product (overlap) between the ground-state
eigenvectors 
\begin{equation}
\label{fidelity}
F(\Gamma,\Gamma + \Delta \Gamma)
  =| \langle \Gamma | \Gamma +\Delta \Gamma \rangle |
              ,
\end{equation}
for proximate interaction parameters, $\Gamma$ and $\Gamma+\Delta \Gamma$,
the so-called fidelity \cite{Uhlmann76,Jozsa94},
provides valuable information as to a distinguishability
of quantum states.
The idea of fidelity also plays a significant role
in the quantum dynamics
\cite{Peres84}
as a measure of tolerance for external disturbances;
see Ref. 
\cite{Gorin06} for a review.

As would be apparent from the definition
(\ref{fidelity}),
the fidelity suits the numerical-exact-diagonalization calculation,
for which 
an explicit expression for 
$|\Gamma\rangle$ is available.
At finite temperatures, the above definition, Eq. (\ref{fidelity}),
has to be modified accordingly, and the modified version of $F$ is
readily calculated
with the quantum Monte Carlo method
\cite{Schwandt09,Albuquerque10,Grandi11}.

Meanwhile,
the fidelity
(\ref{fidelity})
tuned out to be
sensitive to
an onset of criticality
\cite{Quan06,Zanardi06,Qiang08};
see Ref. \cite{Vieira10} for a review.
To be specific,
for a finite-size cluster with $N$ spins,
the fidelity susceptibility
\begin{equation}
\label{susceptibility}
\chi_F = \frac{1}{N} 
      \partial^2_{\Delta \Gamma}F|_{\Delta\Gamma=0}
              ,
\end{equation}
exhibits a notable singularity
at a critical point.
Because
the tractable system size with the numerical-exact-diagonalization method
is restricted severely,
an alternative scheme for criticality
might be desirable to complement
traditional ones.
In fact, the quantum-Monte-Carlo algorithm
applies successfully 
\cite{Albuquerque10}
to the analysis of $\chi_F$ for the $d=2$ transverse-field Ising model
with $N \le 48\times48$ spins.
However, for the frustrated magnetism,
the quantum-Monte-Carlo method suffers from the negative-sign problem.
On the contrary, the numerical diagonalization method
is free from such difficulty,
permitting us to consider a wide range of
intriguing topics.



\begin{figure}
\includegraphics[width=100mm]{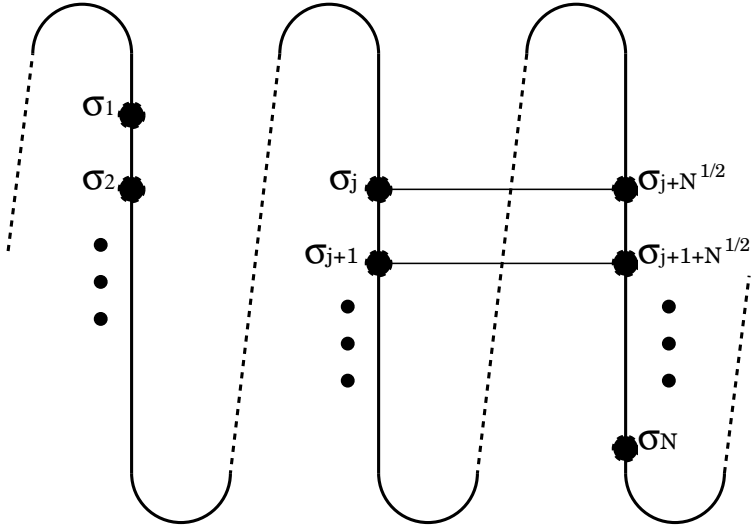}%
\caption{  \label{figure1}
Imposing the screw-boundary condition
\cite{Novotny90,Novotny92},
we 
construct the finite-size
cluster for the two-dimensional
transverse-field Ising model (\ref{Hamiltonian})
with $N$ spins.
As indicated above,
the spins constitute a $d=1$-dimensional
alignment $\{ \sigma_i \}$ ($i=1,2,\dots,N$),
and the dimensionality is lifted to $d=2$ by the bridges (long-range interactions)
over the ($N^{1/2}$)th-neighbor pairs.
The simulation algorithm is explained in Sec. \ref{section2_1}.
}
\end{figure}

In this paper, we calculated the fidelity susceptibility 
(\ref{susceptibility})
for the two-dimensional 
transverse-field (quantum) Ising model 
(\ref{Hamiltonian})
by means of the numerical diagonalization method.
In order to treat a variety of system sizes,
$N=12,14,\dots,32$, systematically,
we implemented
the screw-boundary condition
with the aid of Novotny's method \cite{Novotny90,Novotny92}; see
Fig. \ref{figure1}.
So far, the fidelity susceptibility has been calculated 
\cite{Yu09}
for
$N=10,16,18$, and $20$.
As a comparison, we calculated
the Binder parameter
\cite{Binder81} to determine the location of the critical point.
As a matter of fact,
it has been known that
owing to the screw-boundary condition,
the simulation result suffers from a slowly undulating
deviation with respect to $N$
\cite{Novotny90};
namely, 
for
a quadratic value of $N=16,25$,
the amplitude of deviation gets enhanced.
Such a notorious wavy deviation seems to be suppressed for
$\chi_F$; 
the fidelity susceptibility may serve a promising
candidate for the numerical analysis of critical phenomena.


To be specific,
the Hamiltonian for the two-dimensional transverse-field Ising model
is given by 
\begin{equation}
\label{Hamiltonian}
{\cal H}= -\sum_{\langle ij \rangle} \sigma^z_i \sigma^z_j 
-\Gamma \sum_i^N \sigma^x_i  .
\end{equation}
Here, the Pauli operator $\{  \vec{\sigma}_i \}$ is placed at 
each two-dimensional- (square-) lattice point $i$.
The summation $\sum_{\langle ij\rangle}$ runs
over all possible nearest-neighbor pairs $\langle i j \rangle$.
The parameter $\Gamma$ denotes the transverse magnetic field.
Upon increasing $\Gamma$,
a phase transition between the ferro- and para-magnetic phases
takes place at
$\Gamma_c = 3.04497(19)$ \cite{Hamer00}; see Ref. \cite{Henkel87} as well.
Our aim is to investigate the critical behavior
of 
the fidelity susceptibility \cite{Yu09}
\begin{equation}
\label{exponent}
\chi_F \sim |\Gamma-\Gamma_c|^{-\alpha_F}  ,
\end{equation}
with the critical exponent $\alpha_F$.
As a byproduct,
we calculated the correlation-length critical exponent $\nu$
through resorting to the scaling relation advocated in Ref.
\cite{Albuquerque10};
to avoid a confusion as to the definition of $\nu$,
we refer readers to a brief remark \cite{Gu11}.

The rest of this paper is organized as follows.
In Sec. \ref{section2},
we 
investigate the critical behavior of the fidelity susceptibility
(\ref{susceptibility})
with the numerical diagonalization method;
a brief account for the simulation scheme is given as well.
In Sec. \ref{section3}, we show the summary and discussions.

\section{\label{section2}
Numerical results
}

In this section, 
we present the numerical result for
the transverse-field Ising model 
(\ref{Hamiltonian}).
For the sake of self-consistency,
we give a brief account for the simulation scheme,
namely, Novotny's method
\cite{Novotny90,Novotny92},
to implement the screw-boundary
condition (Fig. \ref{figure1}).
This simulation method 
allows us to
treat a variety of system sizes
$N=12,14,\dots,32$ in a systematic manner.
The linear dimension $L$ of the cluster is given by
\begin{equation}
L=\sqrt{N} ,
\end{equation}
because $N$ spins constitute a rectangular cluster.

\subsection{\label{section2_1}
Simulation method: Screw-boundary condition}

In this section, we explain the simulation scheme
to implement the screw-boundary condition.
Our scheme is based on 
Novotny's method
\cite{Novotny90,Novotny92}, 
which was developed for 
the transfer-matrix simulation of
the classical Ising model.
In order to adapt this method for the quantum-mechanical
counterpart, a slight modification has to be made.
Here, we present a brief, albeit, mathematically closed,
account for the simulation algorithm.

Before commencing an explanation of the technical details,
we sketch a basic idea of Novotny's method.
We consider a finite-size cluster as shown in
Fig. \ref{figure1}.
We place an $S=1/2$ spin (Pauli operator $\vec{\sigma}_i$)
 at each lattice point 
$i(\le N)$.
Basically, the spins 
 constitute a one-dimensional ($d=1$) structure.
The dimensionality is lifted to $d=2$ by the long-range interactions
over the $\sqrt{N}$-th-neighbor distances;
owing to the long-range interaction, the $N$ spins form a
$\sqrt{N}\times \sqrt{N}$
rectangular network effectively.

According to Novotny
\cite{Novotny90,Novotny92}, the long-range interactions
are
introduced systematically
by the use of the
translation operator $P$; see Eq. (\ref{TP_decomposition}).
The operator $P$ satisfies the formula
\begin{equation}
P | \sigma_1,\sigma_2,\dots,\sigma_N \rangle
   = | \sigma_N,\sigma_1,\dots,\sigma_{N-1}\rangle  .
\end{equation}
Here, 
the Hilbert-space bases 
$\{| \sigma_1,\sigma_2,\dots,\sigma_N \rangle \}$
($\sigma_i = \pm 1$) 
diagonalize the longitudinal component
$\sigma^z_i$ of the Pauli operator;
\begin{equation}
\sigma_j^z | \{ \sigma_i \} \rangle  =
\sigma_j   | \{ \sigma_i \} \rangle  .
\end{equation}
The Hamiltonian 
is given by
\begin{equation}
\label{Hamiltonian_novotny}
{\cal H}
= -  
\left[
H \left(1\right)+
H \left(  \sqrt{N} \right)
\right]
-\Gamma \sum_{i=1}^{N} \sigma^x_i
,
\end{equation}
Here, the matrix $H(v)$ denotes 
the
$v$-th neighbor interaction.  
The matrix $H(v)$
is diagonal, and the diagonal element is given by
\begin{equation}
\label{TP_decomposition}
H_{ \{\sigma_i\},\{\sigma_i\} }(v)
=\langle \{\sigma_i\} | H(v) | \{\sigma_i\} \rangle
=\langle \{\sigma_i\} | TP^v | \{\sigma_i\} \rangle
   .
\end{equation}
The insertion of $P^v$ is a key ingredient to introduce
the $v$-th neighbor interaction.
Here, the matrix $T$ denotes
the exchange interaction between
$\{ \sigma_i \}$ and $\{ \tau_i \}$;
namely, the matrix element of $T$ is given by
\begin{equation}
\label{plaquette_interaction}
\langle \{\sigma_i\} |T| \{\tau_i\} \rangle=
\sum_{k=1}^{N}
\sigma_{k}\tau_{k}  
  .
\end{equation}

The above formulae complete the formal basis of our simulation
scheme.
We diagonalize the Hamiltonian matrix
(\ref{Hamiltonian_novotny})
for $N \le 32$ spins numerically.
In the practical numerical calculation,
however,
a number of formulas may be of use;
see
the Appendices of
Refs. \cite{Nishiyama07b,Nishiyama10}.

\subsection{Analysis of the critical point
with
the fidelity susceptibility
$\chi_F$ and Binder's parameter $U$}

In Fig. \ref{figure2},
we present the fidelity susceptivity $\chi_F$
(\ref{susceptibility})
for various $\Gamma$ and $N=12,14,\dots,32$.
Around $\Gamma \approx 3$,
there appears a clear signature of criticality;
in Ref.
\cite{Yu09},
the criticality was analyzed 
for $N=10,16,18$ and $20$.
Our aim is to
survey the critical behavior of $\chi_F$ for 
extended system sizes $N \le 32$.


In Fig. \ref{figure3},
we plot the approximate critical point $\Gamma_c(L)$ 
(plusses)
for $1/L^2(=1/N)$; the range of $N$
 is the same as that 
of Fig. \ref{figure2}.
Here, the approximate critical point denotes the 
location 
of maximal $\chi_F$;
namely,
the relation
\begin{equation}
\label{transition_chi}
\partial_\Gamma \chi|_{\Gamma=\Gamma_c(L)}=0   ,
\end{equation}
holds.
The series of $\Gamma_c(L)$ appears to exhibits a wavy 
(slowly undulating) deviation with respect to $L(=\sqrt{N})$.
Such a wavy character is 
attributed to
an artifact of
the screw-boundary condition \cite{Novotny90};
namely,
the deviation amplitude is suppressed for quadratic values of $N=16,25$
(commensurate condition).
The least-squares fit to the data in Fig.
\ref{figure3}
yields 
an estimate
$\Gamma_c=2.965(46)$ in the thermodynamic
limit.
The result is consistent with 
a preceding estimate $\Gamma_c=2.95(1)$ \cite{Yu09}.
A large-scale numerical-exact-diagonalization result
$\Gamma_c=3.04497(19)$ 
for $N  \le 6 \times 6$
\cite{Hamer00} 
lies out of the error margin.
Possibly,
the abscissa scale $1/L^2$ in Fig. \ref{figure3}, 
namely, the power-law singularity of corrections to scaling,
has to be 
finely-tuned in order to better attain precise extrapolation to the
thermodynamic limit.
Nevertheless, the extrapolated critical point $\Gamma_c$ is no longer
used in the subsequent analyses, and we do not go into further details;
rather, the approximate critical point $\Gamma_c(L)$ 
is fed into the formula, Eq. (\ref{ln_chi_F}).


As a comparison,
we provide an alternative analysis of $\Gamma_c$ 
via the Binder parameter \cite{Binder81}.
In Fig. \ref{figure4},
we present 
the Binder parameter
\begin{equation}
\label{Binder}
U=1-\frac{\langle \Gamma | M^4 | \Gamma \rangle}
{3 \langle \Gamma | M^2 | \Gamma \rangle^2}
   ,
\end{equation}
with
the magnetic moment
\begin{equation}
M= \sum_{i=1}^N \sigma^z_i,
\end{equation}
for various $\Gamma$ and $N=12,14,\dots,32$.
The intersection point of the curves
indicates a
location of criticality.
Because of the above-mentioned wavy deviation,
the location of the intersection point becomes unclear.

In Fig. \ref{figure3},
we plot the approximate critical point $\Gamma_c(L)$ (crosses) for $1/L^2$.
Here, the approximate critical point denotes an intersection point 
of the Binder-parameter curves with
respect
to a pair of system sizes $N=L^2-1$ and $L^2+1$.
Namely, the following relation holds:
\begin{equation}
\label{transition_U}
U(L^2-1)|_{\Gamma=\Gamma_c(L)}=U(L^2+1)_{\Gamma=\Gamma_c(L)}  .
\end{equation}
The finite-size deviation 
of $U$ appears to be
much larger than that of $\chi_F$.
As mentioned above,
such a wavy character is 
attributed to
an artifact of
the screw-boundary condition \cite{Novotny90}.
Namely,
the deviation amplitude gets enhanced for quadratic values of $N=16,25$.
The least-squares fit to these data
yields an estimate $\Gamma_c=3.89(31)$ in the thermodynamic limit $L\to\infty$.
The pronounced finite-size deviation 
prohibits us
from analyzing the criticality reliably.

We address
a remark.
As mentioned above, the oscillatory-deviation amplitude
depends on the condition whether
the system size $L(=\sqrt{N})$ is close to an
integral number (commensurate) or not (incommensurate).
One is able to reduce the oscillatory deviation by tuning the screw pitch
for each $N$ \cite{Nishiyama11}.
Such an elaborate treatment might be worth pursuing 
to better attain precise estimation of critical indices.


\subsection{\label{section2_3}
Fidelity-susceptibility critical exponent $\alpha_F$}

In this section, we analyze the
fidelity-susceptibility critical exponent $\alpha_F$
with the finite-size-scaling method.
As a byproduct, we estimate 
the correlation-length
critical exponent $\nu$.

In Fig. \ref{figure5},
we plot the logarithm of $\chi_F$ at the approximate critical point,
namely, 
\begin{equation}
\label{ln_chi_F}
\ln \chi_F|_{\Gamma=\Gamma_c(L)}   , 
\end{equation}
against $\ln L$ for $N=12,14,\dots,32$
($L=\sqrt{N}$).
According to the finite-size scaling,
at the critical point,
(the singular part of)
the susceptibility obeys the power law  $\chi_F  \sim L^{\alpha_F/\nu}$
with the correlation-length critical exponent $\nu$;
see Ref. \cite{Gu11} as well.
Therefore,
the slope of $\ln L$-$\ln \chi_F|_{\Gamma=\Gamma_c(L)}$ data
indicates the critical exponent
$\alpha_F/\nu$.
The least-squares fit to the data in Fig. \ref{figure5}
yields $\alpha_F/\nu=1.113(49)$.  
This result is to be compared with the preceding one $\alpha_F/\nu=1.02$
\cite{Yu09} for $N=10,16,18$ and $20$.
The series of data in Fig. \ref{figure5}
exhibit a
slowly undulating deviation inherent in
the screw-boundary condition.
As mentioned above, the deviation amplitude depends
on the condition whether the system size
$L(=\sqrt{N})$ is close to an integral number (commensurate) or not (incommensurate).
The system size $12 \le N \le 32$ in Fig. \ref{figure5} covers one period
in the sense that
the difference of the system size $\sqrt{32}-\sqrt{12} \approx 2.2$ 
is close to two (an even number);
owing to the cancellation over one period, 
the result
$\alpha_F/\nu=1.113(49)$ might not be affected by the oscillatory deviation
very much.
As a reference,
we provide
an alternative analysis of $\alpha_F/\nu$.
As mentioned above, 
 the commensurate series
$N=16,24,26$ and the incommensurate one
$N=12,14,30,32$ behave differently.
Among the pairs 
$(N_1,N_2)$
($L_{1,2}=\sqrt{N_{1,2}}$)
within each series,
we 
calculate the exponent
\begin{equation}
\alpha_F/\nu=
\frac{\ln \chi_F(L_1)|_{\Gamma=\Gamma_c(L_1)}
-\ln \chi_F(L_2)|_{\Gamma=\Gamma_c(L_2)}
}{
\ln L_1-\ln L_2
}
                .
\end{equation}
For the commensurate-series pairs, $(N_1,N_2)=(16,24)$ 
and $(16,26)$,
we arrive at $\alpha_F/\nu=1.126$ and
$1.068$, respectively.
Similarly, for the incommensurate-series pairs,
$(12,14)$,
$(30,32)$,
$(12,30)$,
$(14,32)$, and
$(12,32)$,
we obtain 
$\alpha_F/\nu=1.191$,
$1.157$,
$1.133$,
$1.124$, and
$1.135$, respectively.
The least-squares fit to these data with the abscissa scale
$1/(N_1+N_2)^2$ yields 
$\alpha_F/\nu=1.10(3)$ 
in the thermodynamic  limit $N\to\infty$.
This result confirms the
above preliminary result
$\alpha_F/\nu=1.113(49)$; hereafter,
we accept $\alpha_F / \nu=1.113(49)$,
aiming to estimate related scaling indices.
Nevertheless, we stress that 
a rather moderate cluster size already reaches the scaling
regime, as noted in Ref. \cite{Yu09}.


We are able to estimate the respective indices, $\alpha_F$ and $\nu$,
through resorting to a number of scaling relations.
According to the scaling argument \cite{Albuquerque10},
the index $\alpha_F$ satisfies the
relation 
\begin{equation}
\label{scaling_relation}
\alpha_F = \alpha+\nu
  ,
\end{equation}
 with the specific-heat critical exponent
$\alpha$.
On the one hand, the 
hyper-scaling theory insists that the specific-heat 
critical exponent satisfies the relation
$\alpha=2-D\nu$ with the spatial and temporal dimensionality $D=2+1$.
Putting the present estimate 
$\alpha_F / \nu =1.113(49)$ into the above scaling relations,
we arrive at
$\alpha_F=0.715(20)$ and   
$\nu=0.642(10)$.
The former is in good agreement with
the preceding estimate $\alpha_F=0.73$ \cite{Yu09} for
$N=10,16,18$ and $20$.
The latter lies slightly out of 
a recent Monte Carlo result
$\nu=0.63020(12)$
 \cite{Deng03} for the three-dimensional classical Ising model;
the universality class of
the ground-state phase transition for the two-dimensional transverse-field Ising model
belongs to the three-dimensional classical counterpart.
Again, it is suggested that
the $\chi_F$-based finite-size scaling analysis is
less influenced by corrections to scaling.
In particular, an agreement with $\alpha_F=0.73$ \cite{Yu09}
confirms that a rather moderate system size  $N \approx 4\times4$
already reaches the scaling regime.

\section{\label{section3}
Summary and discussions}

The critical behavior of the fidelity susceptibility 
(\ref{susceptibility})
was investigated for the two-dimensional transverse-field Ising model 
(\ref{Hamiltonian})
by means of the numerical diagonalization method.
In order to treat a variety of system sizes $N=12,14,\dots,32$,
we implemented
the screw-boundary condition 
(Sec. \ref{section2_1})
with the aid of Novotny's method \cite{Novotny90,Novotny92}.

\begin{figure}
\includegraphics[width=100mm]{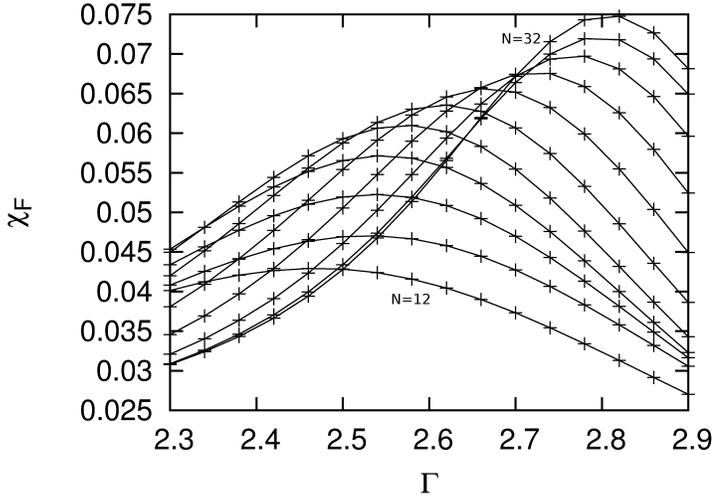}%
\caption{  \label{figure2}
The fidelity susceptibility $\chi_F$ 
(\ref{susceptibility}) 
is plotted for various $\Gamma$ and $N=12,14,\dots,32$.
A notable singularity emerges at $\Gamma \approx 3$.
The drift of $\Gamma_c$ is analyzed in 
Fig. \ref{figure3}.
}
\end{figure}

\begin{figure}
\includegraphics[width=100mm]{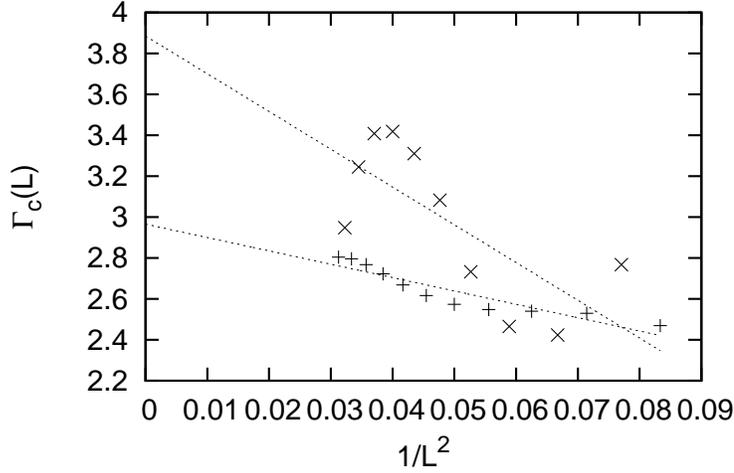}%
\caption{  \label{figure3}
The approximate critical point $\Gamma_c(L)$
determined via $\chi_F$ (plusses)
[Eq. (\ref{transition_chi})]
and $U$ (crosses)
[Eq. (\ref{transition_U})]
is plotted for $1/L^2$.
The least-squares fit to these data yields
$\Gamma_c=2.965(46)$ and $3.89(31)$, respectively,
in the thermodynamic limit $L\to\infty$. 
The series of data exhibit a slowly undulating deviation
intrinsic to
the screw-boundary condition \cite{Novotny90};
that is,
for quadratic values of $N=16$ and $25$,
the 
deviation amplitude gets enhanced.
}
\end{figure}

\begin{figure}
\includegraphics[width=100mm]{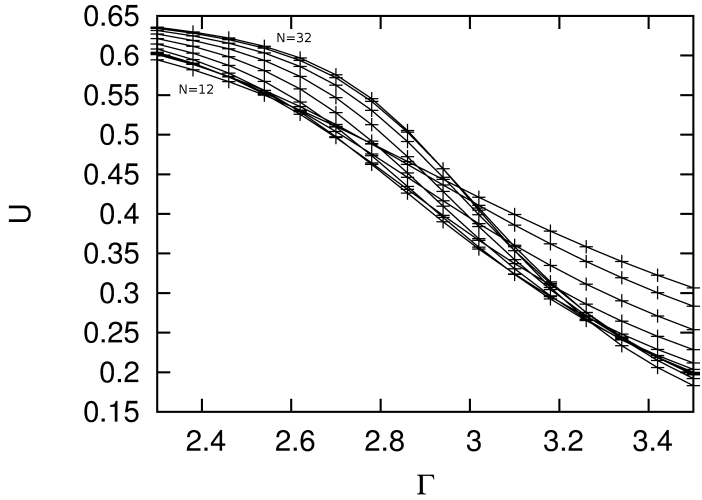}%
\caption{  \label{figure4}
The Binder parameter $U$ 
(\ref{Binder})
is plotted for various $\Gamma$ and $N=12,14,\dots,32$.
The intersection point of the curves 
indicates a location of the
critical point $\Gamma_c(L)$
(\ref{transition_U}), which is
shown in Fig. \ref{figure3}.
}
\end{figure}

\begin{figure}
\includegraphics[width=100mm]{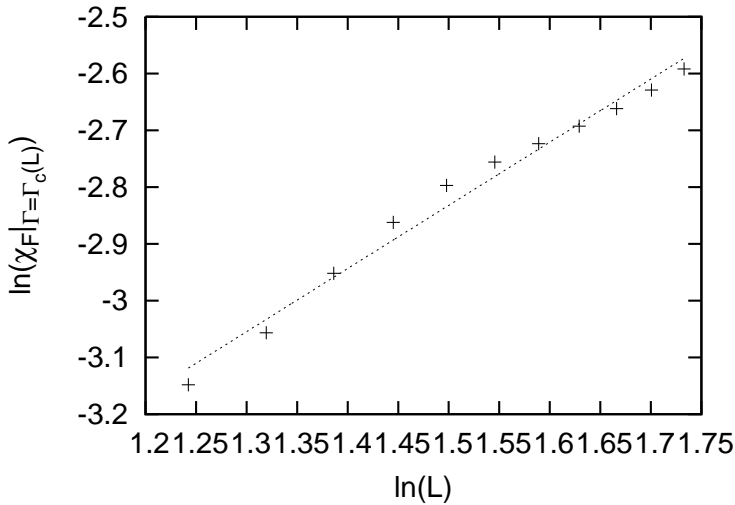}%
\caption{  \label{figure5}
The logarithm of the fidelity susceptibility at the 
approximate critical point,
namely,
$\ln \chi_F|_{\Gamma=\Gamma_c(L)}$,
is plotted against $\ln L$
for $N=12,14,\dots,32$ ($L=\sqrt{N}$).
The slope of the series of data indicates 
the critical exponent $\alpha_F/\nu$.
The least-squares fit to these data yields
an estimate
$\alpha_F/\nu=1.113(49)$.
A
further consideration of
the respective indices, $\alpha_F$ and $\nu$, is made
in the text.
}
\end{figure}

The fidelity susceptibility exhibits a notable singularity
(Fig. \ref{figure2}), with which we estimated
the critical point 
as $\Gamma_c=2.965(46)$ (Fig. \ref{figure3}).
Moreover, 
scrutinizing 
its power-law singularity at the
critical point (Fig. \ref{figure5}),
we obtained the
critical exponent 
$\alpha_F=0.715(20)$.
These results are to be compared with
the preceding estimates \cite{Yu09},
$\Gamma_c=2.95(1)$ and $\alpha_F=0.73$,
obtained for $N=10,16,18$ and $20$.
Hence, it is suggested that
the simulation results
for rather small clusters already reach the scaling regime.
As a byproduct, through the scaling relation
(\ref{scaling_relation}) \cite{Albuquerque10},
we arrive at 
$\nu=0.642(10)$.
According to an extensive Monte Carlo
simulation for the three-dimensional
classical
Ising model \cite{Deng03}, the 
correlation-length critical exponent was estimated as
$\nu=0.63020(12)$; this value lies slightly 
out of the error margin.
Again, it is suggested that the analysis via $\chi_F$ might be less
influenced by corrections to scaling (systematic errors).
As mentioned in Introduction,
the quantum-Monte-Carlo method is also a clue to the analysis
of the fidelity susceptibility.
Actually,
a considerably precise result $\nu=0.625(3)$
was reported in Ref. \cite{Albuquerque10}.



By definition (\ref{fidelity}), 
the fidelity susceptivity suits the numerical diagonalization
calculation.
Because the tractable system size with the numerical
diagonalization method is severely restricted,
such 
an alternative scheme might be 
desirable to complement the existing ones.
It would be tempting to apply the fidelity susceptibility to 
a wide class of systems of current interest
such as
the frustrated quantum magnetism, for which the Monte Carlo method
suffers from the negative-sign problem.
This problem would be addressed in the future presentation.


\bibliographystyle{elsarticle-num}







\end{document}